# Mode-locked laser timing jitter limitation in optically enabled, spectrally sliced ADCs

Andrea Zazzi
*Institute of Integrated Photonics*
*RWTH Aachen University*
Aachen, Germany
azazzi@iph.rwth-aachen.de

Juliana Müller
*Institute of Integrated Photonics*
*RWTH Aachen University*
Aachen, Germany
jmueller@iph.rwth-aachen.de

Sergiy Gudyriev
*Heinz Nixdorf Institute*
*Paderborn University*
Paderborn, Germany
sergiy.gudyriev@uni-paderborn.de

Pablo Marin-Palomo
*Institute of Photonics and Quantum Electronics*
*Karlsruhe Institute of Technology*
Karlsruhe, Germany
pablo.marin@kit.edu

Dengyang Fang
*Institute of Photonics and Quantum Electronics*
*Karlsruhe Institute of Technology*
Karlsruhe, Germany
dengyang.fang@kit.edu

J. Christoph Scheytt
*Heinz Nixdorf Institute*
*Paderborn University*
Paderborn, Germany
christoph.scheytt@hni.uni-paderborn.de

Christian Koos
*Institute of Photonics and Quantum Electronics*
*Karlsruhe Institute of Technology*
Karlsruhe, Germany
christian.koos@kit.edu

Jeremy Witzens
*Institute of Integrated Photonics*
*RWTH Aachen University*
Aachen 52074, Germany
jwitzens@iph.rwth-aachen.de

*Abstract*—Novel analog-to-digital converter (ADC) architectures are motivated by the demand for rising sampling rates and effective number of bits (ENOB). The main limitation on ENOB in purely electrical ADCs lies in the relatively high jitter of oscillators, in the order of a few tens of fs for state-of-the-art components. When compared to the extremely low jitter obtained with best-in-class Ti:sapphire mode-locked lasers (MLL), in the attosecond range, it is apparent that a mixed electrical-optical architecture could significantly improve the converters' ENOB. We model and analyze the ENOB limitations arising from optical sources in optically enabled, spectrally sliced ADCs, after discussing the system architecture and implementation details. The phase noise of the optical carrier, serving for electro-optic signal transduction, is shown not to propagate to the reconstructed digitized signal and therefore not to represent a fundamental limit. The optical phase noise of the MLL used to generate reference tones for individual slices also does not fundamentally impact the converted signal, so long as it remains correlated among all the comb lines. On the other hand, the timing jitter of the MLL, as also reflected in its RF linewidth, is fundamentally limiting the ADC performance, since it is directly mapped as jitter to the converted signal. The hybrid nature of a photonically enabled, spectrally sliced ADC implies the utilization of a number of reduced bandwidth electrical ADCs to convert parallel slices, resulting in the propagation of jitter from the electrical oscillator supplying their clock. Due to the reduced sampling rate of the electrical ADCs, as compared to the overall system, the overall noise performance of the presented architecture is substantially improved with respect to a fully electrical ADC.

## I. Introduction

The growing demand for high sampling rate data converters poses difficult engineering challenges given the limitations in bandwidth and resolution of state-of-the-art electronic data converters, with the latter mainly originating from the aperture jitter of electronic clocks [1]. To overcome such limitations, different approaches have been taken, e.g., replacing quartz crystals with dielectric sapphire resonators operating at low temperature, which feature much improved phase noise levels [2]. Another approach consists in using optical-electrical hybrid integration, exploiting the attosecond jitter of low-jitter optical pulse trains as provided by Ti:sapphire lasers [3] or Er-doped fiber MLLs [4]. Together with the broadband signal processing capability of electro-optical systems, this enables record-high ENOB high-speed photonic data converters [5].

In the following, we analyze the optically enabled, spectrally sliced ADC whose architecture is shown in Fig. 1, focusing on the evaluation of the fundamental ENOB limitations arising from the optical noise sources. In this spectrally sliced ADC, the RF signal that is to be digitized is first amplitude modulated onto a single-frequency optical carrier by means of a broadband electro-optical modulator implemented as a Mach-Zehnder modulator (MZM) biased at its quadrature point. The resulting optical signal is subsequently sliced into multiple spectral slices, each with a reduced bandwidth, using higher-order optical filters implemented as coupled-resonators optical waveguide (CROW) filters. Each slice is then mixed by means of a 3-dB directional coupler/splitter (DCS) with a reference tone provided by an MLL, whose repetition rate is matched to the slices' bandwidth, allowing the down-conversion of each slice to baseband electrical signals. MLL comb lines are chosen to be to the side of the analyzed spectrum, with a small additional guard-band, resulting in heterodyne coherent detection. Single channels are detected by an array of balanced photodetector (PD) pairs, digitized by standard all-electrical ADCs, and finally stitched back together by digital signal processing (DSP). An analytical description of the system is given in the next section.

The goal of this paper is the evaluation and quantification, in terms of the converter's signal-to-noise-ratio (SNR) and ENOB, of the noise arising from the additional optical sources employed in the system, including phase noise and corresponding timing jitter. The effects of the single frequency laser phase noise, as well as of the MLL phase noise and timing jitter, are modeled in details. Section II analytically describes the system model. The resulting SNR is analytically derived in Section III and numerically verified in Section IV. Section V finally applies the derived models to predict the performance achievable in practical optically enabled, spectrally sliced ADCs given the performance of currently available electronics and light sources.

The authors acknowledge funding from the Deutsche Forschungsgemeinschaft (DFG) for project "Ultra-Wideband Photonically Assisted Analog-to-Digital Converters" (PACE, no. 650602) and from the European Research Council (TeraSHAPE, no. 773248).

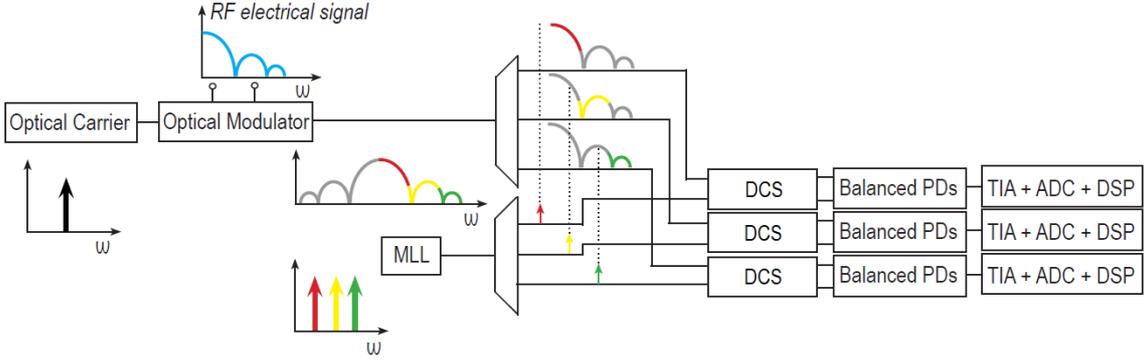

Fig. 1. Optically enabled, spectrally sliced ADC architecture. Spectral slices are individually analyzed by coherent receivers supplied with reference tones by an MLL. The resulting information is aggregated in the digital domain to recover the broadband signal.

## II. System Model

First, the output of the MLL serving as a spectral reference is modeled, following the derivation in [6], as a sum of $N$ individual optical lines with locked phases as a consequence of four-wave-mixing (FWM)

$$E_{MLL}(t) = e^{j[\omega_c t + \theta_c(t)]}\left[\sum_{n=1}^{N} E_n e^{j[(n-n_c)\omega_r(\Delta t_r(t)+t)+\phi_n]}\right] + c.c. \quad (1)$$

Equivalently, this corresponds to a carrier with angular frequency $\omega_c$ and optical phase noise $\theta_c(t)$ modulated by a pulse train with angular repetition frequency $\omega_r$ and timing jitter $\Delta t_r(t)$. $n_c$ represents the index of the central comb line and $\phi_n$ are static phase offsets resulting from dispersion in the system. The real valued amplitude of individual comb lines $E_n$ is considered constant, neglecting relative intensity noise (RIN), since it is usually dominated by phase noise [7].

Next, the broadband electro-optical modulator, which transduces into the optical domain the RF signal that is to be digitized, is described. The generated optical signal, at the output of the modulator, is cut into multiple spectral slices by means of optical filters, each with a passband $\omega_r$ matched to the free spectral range (FSR) of the MLL. The electric field of each slice can be described as

$$E_{MZM,m}(t) = e^{j(\omega_0 + m\cdot\omega_r)t + j\theta_0(t)}\int_{\omega_s = m\cdot\omega_r - \omega_{\overline{\Delta}}}^{\omega_s = (m+1)\cdot\omega_r - \omega_{\overline{\Delta}}} E_s(\omega_s)e^{j(\omega_s - m\cdot\omega_r)t}d\omega_s + c.c. \quad (2)$$

where $m$ is the slice index, $\omega_0$ the angular frequency of the optical carrier and $\theta_0(t)$ its phase noise, and $\omega_s$ the offset frequency from the carrier. $E_s(\omega_s)$ represents the time-dependent transfer function of the modulator, determined by the applied electrical signal. Only the carrier and either the upper or lower signal sidebands are needed to reconstruct the signal, so that the best spectral efficiency is obtained e.g. when the carrier coincides with the lower edge of the lowest slice, and the slices are used to record the upper sidebands. $\omega_{\overline{\Delta}}$ is a small guard-band used to reduce the truncation of the carrier's spectrum due to filtering, which has a finite linewidth in the presence of phase noise, and corresponds to a frequency offset by which the carrier is shifted into the first slice.

In order to digitize the signal, each slice is mixed with an MLL line by means of an optical coupler/splitter (DCS) prior to photodetection, down-converting the slice to a baseband signal. MLL lines are singulated from the original comb using optical filters and are offset by $\omega_\Delta$ relative to the lower edge of the signal slice. The mixed signal, at the output of the DCS, is then detected by a pair of balanced photodetectors (PD). The second guard-band $\omega_\Delta$ ensures that all the beat note frequencies have the same sign, so that the implementation of a 90-degree hybrid is not required, and pushes up the lowest beat note frequency that needs to be detected, so that electronic flicker noise arising from the transimpedance amplifiers (TIA) [8] can be reduced.

Coherent detection, with the reference tone in the center of each slice, i.e., intradyne detection, would allow dual-sided spectrum detection, meaning that each slice would have twice the effective bandwidth. However, this would also require a 90-degree hybrid as well as two balanced PD pairs and differential TIAs per channel. The heterodyne detection scheme opted for here thus results in the same overall bandwidth for a given number of TIAs, only requires a DCS instead of a more complex 90-degree hybrid and is compatible with AC-coupling of electrical signals without loss of information.

The photocurrent at the output of the balanced PD-pairs, for the $m^{th}$ slice, can be then expressed as

$$i_m = \frac{1}{2}|E_{MZM,m}(t) + E_{MLL,m}(t)|^2 - \frac{1}{2}|E_{MZM,m}(t) - E_{MLL,m}(t)|^2 = \quad (3)$$

$$2e^{-j(\omega_\Delta + \omega_{\overline{\Delta}})t + j(\theta_0(t)-\theta_c(t))}E_{n_m}^* e^{-j[(n_m - n_c)\omega_r \Delta t_r(t) + \phi_{n_m}]} \cdot \int_{\omega_s = m\cdot\omega_r - \omega_{\overline{\Delta}}}^{\omega_s = (m+1)\cdot\omega_r - \omega_{\overline{\Delta}}} E_s(\omega_s)e^{j(\omega_s - m\cdot\omega_r)t}d\omega_s + c.c.$$

Photocurrents are subsequently digitized by means of an array of electric ADCs and sent for aggregation to a DSP, stitching together data from different slices. The harmonic conjugate of the photocurrent is computed with the Hilbert transform $H(i_m)$ and the complex valued phasor representation of the photocurrent generated by adding the two components in quadrature. Since all signal components lie above the reference tones provided by the comb lines, the beat notes can be unambiguously mapped. Finally, each phasor is upconverted to the corresponding original frequency band by digitally shifting it up in frequency by $m\omega_r$ and the static phase error $\phi_{n_m}$ corrected by one-tap equalization. The reconstructed signal can then be written as

$$S = \left|\sum_m \frac{i_m + j\cdot H(i_m)}{E_{n_m}}e^{jm\omega_r t + j\phi_{n_m}}\right|^2 = \quad (4)$$

$$16\left|\sum_m e^{-jm\omega_r \Delta t_r(t)}\int_{\omega_s = m\cdot\omega_r - \omega_{\overline{\Delta}}}^{\omega_s = (m+1)\cdot\omega_r - \omega_{\overline{\Delta}}} E_s(\omega_s)e^{j\omega_s t}d\omega_s\right|^2$$

A number of techniques can be used to estimate the static phase errors $\phi_{n_m}$ so that they can be corrected. Most straightforwardly, sinus pilot tones can be sent through the system and the phase error of the digitized signal extracted – which directly reflects the phase error corresponding to the slice in which the optical signal's sideband is recorded. Alternatively, the phase error can also be estimated as part of the spectral stitching algorithm [9].

It can be seen that if the MLL is assumed to be free of timing jitter, i.e., $\Delta t_r = 0$, the signal is perfectly reconstructed. On the other hand, the presence of MLL timing jitter clearly degrades signal reconstruction, as will be further analyzed in more details in the following sections.

## III. ANALYTICAL SNR

In this section, the SNR of an optically enabled, spectrally sliced converter is analytically derived. The SNR of a standard electrical ADC with aperture jitter $\Delta t$ and at angular analog input signal frequency $\omega$ is given by [10]

$$\text{SNR} = 20\log_{10}\left[\frac{1}{\omega \cdot \Delta t}\right] \quad (5)$$

In Eq. (4), signal components generated by the $m^{\text{th}}$ slice feature a phase noise $m\omega_r \Delta t_r(t)$ instead of $\omega_{RF}\Delta t_r(t)$ as expected from a one-to-one propagation of the jitter from the MLL to the digitized RF signal, where $\omega_{RF}$ is its frequency. The effective MLL timing jitter seen at the system output can thus be readily seen as being rescaled as $(m\omega_r/\omega_{RF})\Delta t_r(t)$. In other words, the signal frequency applied to Eq. (5) is rounded down to the next lower multiple of MLL pulse repetition frequencies, resp. slice widths.

The signal frequency seen by the electrical ADC in slice $m$ is the residual $\omega_{RF} - m\omega_r$, and the resulting phase noise $(\omega_{RF} - m\omega_r) \cdot \Delta t_e(t)$, so that the equivalent of Eq. (5) for the optically enabled, spectrally sliced ADC is

$$\text{SNR} = 20\log_{10}\left[\frac{1}{\sqrt{\left[m\omega_r \cdot \Delta t_r(t)\right]^2 + \left[(\omega_{RF} - m\omega_r) \cdot \Delta t_e(t)\right]^2}}\right] \quad (6)$$

with $\Delta t_e(t)$ the timing jitter of the electrical oscillator supplying the clock to the electrical ADCs. As the two random processes are fully uncorrelated, the corresponding phase errors are added as a sum of variances.

While Eq. (6) does not require specific assumptions in regards to the spectral shape of the oscillators' phase noise, in Section IV it is numerically validated assuming oscillator phase noise to follow Wiener processes. The electrical oscillator linewidth $\Delta\omega_e$ and the MLL beat-note linewidth $\Delta\omega_r$ (its RF linewidth) can then be related to the electric clock and MLL pulse train jitter via [11]

$$\Delta t_{e/r}(t) = \frac{1}{\omega_{e/r}}\sqrt{\Delta\omega_{e/r} t} \quad (7)$$

where $\omega_e$ is the angular frequency of the RF oscillator.

As expressed by Eq. (6), the SNR is limited by both the MLL jitter and the electrical oscillator jitter, whose variances are respectively weighted by $(m\omega_r)^2$ and $(\omega_{RF} - m\omega_r)^2$. While the electrical jitter still plays a role, it is much reduced for high-speed signals falling in higher order slices and the overall SNR is improved provided $\Delta t_r(t) < \Delta t_e(t)$.

It is noteworthy that to the first order the SNR does not depend on the phase noise of either the optical carrier sent through the modulator, $\theta_0(t)$, or the correlated phase noise of the MLL, $\theta_c(t)$. As will be discussed in more details in Section IV, these matter to some extent. However, their effect on the SNR can be reduced with adequate signal processing, while the limit expressed by Eq. (6) is fundamental. As already mentioned above, the carrier of the modulated signal may be partially filtered out if the guard-band $\omega_{\bar{\Delta}}$ is not large enough. This can lead to a reduction of the SNR, as numerically exemplified in Section IV. Moreover, isolating a single comb line from the MLL to serve as a reference tone for a given slice also results in a truncation of its spectrum. This partially converts phase noise into RIN, that propagates to the digital output in the form of a signal-amplitude dependent noise, thus also degrading the SNR. However, this can be straightforwardly compensated by monitoring the instantaneous optical power of each line, $|E_n(t)|^2$, after optical filtering, for example by means of an optical power tap. The time-varying value of $E_n(t)$ can then be normalized out during signal reconstruction as performed in Eq. (4).

## IV. NUMERICAL VALIDATION

A compact numerical model has been developed to validate the analytical results derived in the previous section. The considered noise sources are the phase noise of the modulated carrier, $\theta_0(t)$, the phase noise correlated among all the MLL comb lines, $\theta_c(t)$, the uncorrelated MLL phase noise resulting from its pulse train jitter $\Delta t_r(t)$, and the jitter of the electrical oscillator supplying the clock to the electrical ADCs used in the individual slices, $\Delta t_e(t)$. As described in Section III, $\Delta t_r(t)$ and $\Delta t_e(t)$ are respectively associated to the linewidths $\Delta\omega_r$ and $\Delta\omega_e$. These four phase noise terms are modeled as independent Wiener processes.

We model a system composed of 4 slices, each with a 30 GHz bandwidth matched to the FSR of the MLL, providing a total bandwidth of 120 GHz. The assumptions made for the numerical evaluation were constrained by the code's memory requirements. To prevent aliasing, the simulation time step is set to 0.3 ps. To be able to study the effect of phase noise, the total simulation time has to be long enough for sufficient timing jitter to accumulate, as given by Eq. (7). A value of 3.3 μs was chosen, already consisting in over $10^7$ time-steps. This constrained the smallest linewidth values that could be assumed while still obtaining meaningful results. Given these constraints, the linewidth of the modulated optical carrier was chosen to be 100 kHz, as typical for a telecom-grade external cavity laser (ECL). The MLL, on the other hand, was modeled as having an optical linewidth of 10 MHz and an RF linewidth of 3 kHz, which are not only very significantly above the linewidths of state-of-the-art Ti:sapphire or Er-doped fiber lasers [3],[4], but also corresponds to a pulse timing jitter much larger than the timing jitter of an electrical oscillator. In the following simulations, the linewidth of the electrical oscillator was also exaggerated and assumed to be 180 kHz, which is many orders of magnitude higher than the linewidth of an oven-controlled quartz crystal oscillator (OCXO) upconverted to the required 60 GHz clock-rate for the 30 GS/s electrical ADCs, but is exaggerated by a similar factor as the MLL's RF linewidth if compared to a best-in-class Er-doped fiber laser [4]. In the following, the effect of individual noise terms is evaluated one by one, so that the RF linewidth of the MLL and the linewidth of the electrical oscillator can be straightforwardly rescaled to realistic values, as done in Section V.

The effects of each noise source on the reconstructed signal are first individually analyzed using an exemplary 100 GHz sinusoidal signal falling into the fourth slice. The model for the overall performance evaluation is then verified over the whole 120 GHz frequency range of the spectrally sliced ADC.

First, solely the phase noise of the modulated carrier is considered. All the other noise sources are set to zero. In Fig. 2(a), the reconstructed signal, plotted as a function of time in red, is overlaid with the input signal plotted in blue (wherein the reconstructed signal has been doubled to compensate for the spectrally sliced ADC only analyzing the upper optical sidebands). The reconstructed signal is plotted as a function of the input signal in Fig. 3(a).

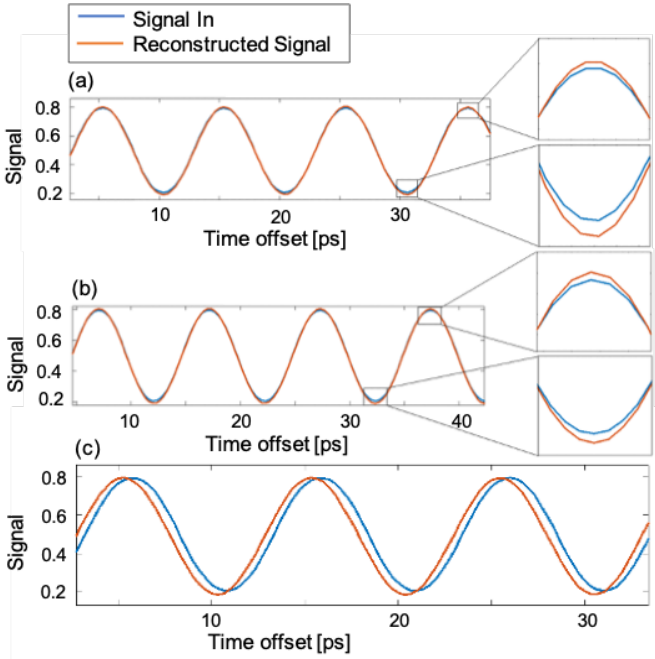

Fig. 2. Evolution in time of the 100 GHz signal at the output of the modulator (blue) as well as of the reconstructed signal (red), after digitization with four slices. The amplitude of the reconstructed signal has been doubled in these graphs to compensate for only one sideband being analyzed. (a) Only carrier phase noise is considered. (b) Only correlated MLL phase noise is considered. The output noise in (b) and (c) takes the form of a time varying deviation of the reconstructed signal amplitude with respect to the input reference signal and cannot be categorized as apperture jitter. (c) Only pulse train timing jitter is considered. The noise can be seen to take the form of a time offset between the reconstructed and the input signals, that slowly varies as further RF phase noise is being accumulated, representing the sampling-time uncertainty.

Some level of distortion is noticeable, that results from the spectrally sliced ADC only considering the upper optical sidebands. This is easily verified by comparing the reconstructed signal to a version of the optical input signal fed through an ideal, 120 GHz wide optical filter rejecting the lower sidebands, as shown in Fig. 3(b). It can be seen that the reconstruction is now perfect and noiseless, confirming that distortion and noise seen in Fig. 3(a) are only due to the truncation of the signal in the frequency domain. Distortion could be prevented by recording both the upper and the lower sidebands with the spectrally sliced ADC, or, more elegantly since it does not require doubling the amount of hardware, reconstruct the lower sidebands based on the a-priori knowledge that the optical signal has been generated by an MZM biased at its 3-dB point.

The noise observed in Fig. 3(a) is due to a partial truncation of the carrier of the modulated signal. To maximize the overall system bandwidth, one would align the optical carrier with the lower edge of the lowest slice, so that the full system bandwidth would be available for sideband analysis. In order to reduce the noise to the level observed in Fig. 3(a), we have pushed the optical carrier by a small amount $\omega_{\overline{\Delta}}$ into the lowest slice, so as to reduce truncation of its spectrum. This guard-band was set to 2 GHz and is on the order of what would be required in a practical system since a realistic carrier linewidth has been chosen in the simulations.

Next, the effect of the correlated phase noise, shared among all the MLL comb lines and corresponding to its optical linewidth, has been investigated. Once more, a 100 GHz sinusoidal signal is fed to our system, switching off all the other noise sources. By overlaying the reconstructed signal with the input signal (Fig. 2(b)), as previously, we again observe amplitude noise in the reconstructed signal, that can be further seen to be signal level dependent in Fig. 3(c). It originates from RIN in the reference tones provided by the filtered MLL comb lines, as previously described. As already discussed in Section III, it is in principle possible to cancel this noise by measuring the instantaneous optical power of the filtered comb lines and use this to normalize the signal components during signal reconstruction, as expressed by Eq. (4). Fig. 3(d) shows the reconstructed signal, assuming digital cancelation of the RIN, plotted against the input signal. Noiseless reconstruction is achieved, confirming the above.

Finally, the effect of MLL timing jitter has been assessed. Once again, a 100 GHz sinusoidal input signal is considered, with all the other noise sources turned off. Comparing the original signal with the digitized output, with the overlay of the two signals shown in Fig. 2(c), it is apparent that the effect of noise is much more pronounced in this scenario, even though the RF linewidth of the MLL was by far the lowest linewidth considered. Moreover, the two signals can be seen to be time-shifted relative to each other, with a time offset that varies over time, as expected in the case of aperture jitter noise. This is further confirmed in Fig. 3(e), plotting the reconstructed signal as a function of the input signal, in which the noise levels are highest at mid-input-signal-levels, where the input signal has the highest derivative.

This digitized signal distortion cannot be compensated by digital post-processing or simple architectural modifications, as done for the other noise sources, and represents a fundamental performance limitation of the optically enabled ADC.

In a complete electro-optical noise model, the aperture jitter of the electrical oscillator supplying the shared clock for the electrical ADC array also limits the system performance.

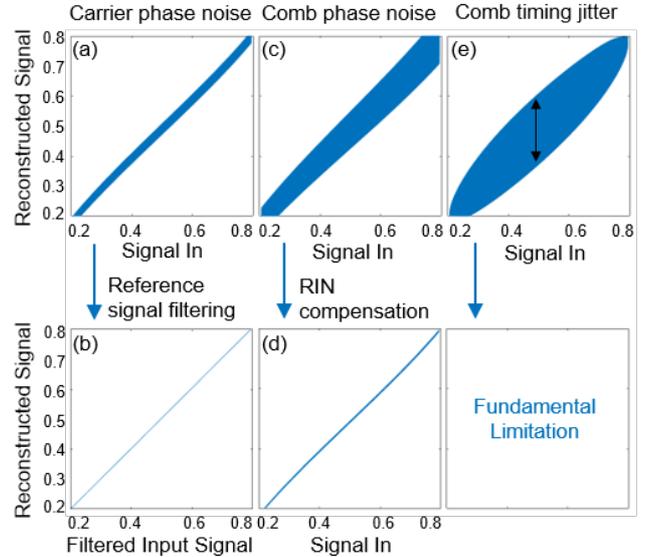

Fig. 3. Summary of the impact of different categories of noise on the digitization of a 100 GHz input signal employing four 30 GHz slices. The reconstructed signal is plotted against a signal reference. In (a) and (b) only carrier phase noise is considered. In (a) the reconstructed signal is plotted against the signal at the outut of the modulator, while in (b) it is plotted against a reference signal generated by filtering the signal at the output of the modulator with an ideal 120 GHz filter. The filtered input signal has also been doubled to compensate the loss of the lower sidebands in the overall signal strength. Distortion and noise are seen in (a), while (b) shows perfect reconstruction of the modified reference signal. In (c) and (d), only the correlated phase noise shared among all the MLL lines is considered. In (c), the reconstructed signal is affected by RIN resulting from spectral truncation of the comb lines serving as reference tones. In (d), this RIN is canceled during digital signal processing and noiseless signal reconstruction is recovered. In (e), only the uncorrelated MLL phase noise resulting from pulse train jitter is considered. The observed noise levels in the recovered signal are proportional to the derivative of the input signal.

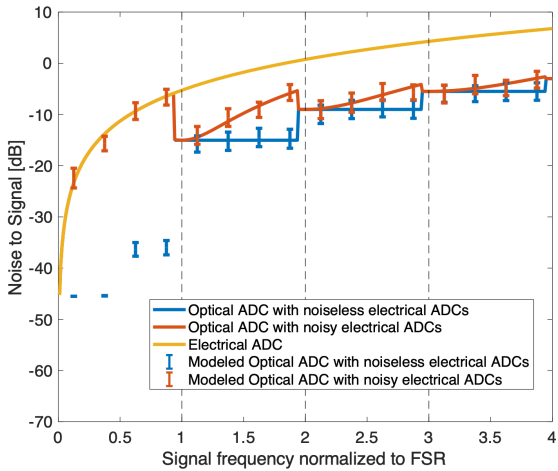

Fig. 4. Noise-to-signal ratio of the optically enabled ADC as a function of the RF input signal frequency. Solid lines correspond to Eq. (6) in the presence of electric oscillator phase noise only (yellow, for an all-electric ADC), MLL jitter only (blue), or a combination of both (red). Theoretical estimates are overlaid by simulation results with error bars showing the $3\sigma$ confidence intervals.

For comparison, the SNR of a fully electrical ADC, as calculated with Eq. (5), is plotted in Fig. 4 with the yellow line. The SNR of the analyzed optically enabled, spectrally sliced ADC is shown in blue for the scenario with MLL pulse train jitter as the only noise source. In this case, the system features a constant SNR for signal frequencies falling within one spectral slice, since the frequency applied to the MLL timing jitter is rounded down to the next lower integer number of slice widths. Finally, the SNR calculated with Eq. (6) considering both jitter sources is plotted in red. In this scenario, the performance of the converter shows a net improvement with respect to the fully electrical solution, as a consequence of the MLL pulse train jitter $\Delta t_r(t)$ being assumed to be much smaller than the electrical jitter $\Delta t_e(t)$. Curves plotted as solid lines are analytically calculated according to Eq. (6), as derived in Section III. They are overlaid by the results of the numerical simulations represented by error bars showing the $3\sigma$ confidence intervals estimated with 65 simulations for each data point. The excellent overlay validates the analytical model.

## V. DISCUSSION AND CONCLUSION

The results reported in the previous section are based on assumptions constrained by the simulation method and serve to validate Eq. (6), that can be easily applied to the noise characteristics of a meaningful practical implementation. The main motivation of this work is the exploitation of ultra-low jitter MLLs, such as Er-doped fiber lasers, that offer timing jitter levels as low as 870 attoseconds when integrated in the 100 Hz to 10 MHz range [4]. State-of-the-art OCXOs, as required to supply the clock of the electrical ADCs used in the individual slices, have a jitter in the order of 6.4 fs when integrated over the same frequency range [12].

In a four-slice configuration, as explored here, the effective aperture jitter of the electrical oscillator is reduced by a factor four, since the maximum electrical frequency is the system bandwidth divided by the number of slices. The effective MLL pulse train jitter is rescaled by a factor $(M-1)/M$, with $M$ the number of slices, that is equal to $3/4$ here. The effective jitter values are then added as a sum of squares (sum of variances), according to Eq. (6), since they are independent from each other. The resulting 1.6 fs effective electric oscillator jitter remains substantially larger than the 652 attosecond effective MLL jitter and remains the primary performance limitation at four slices (as the number of slices is increased, the performance limit asymptotically reaches the value determined by the MLL pulse train jitter only, as given by Eq. (5)). The resulting SNR can be converted to an ENOB of 9.4, 2 bits better than for an all-electric ADC given the OCXO jitter.

These numbers may seem very high for a 120 GHz ADC, particularly as the commercial state-of-the-art ENOB of all-electrical ADCs is on the order of 5 at such speeds, limited by aperture jitter in the order of 20 fs [13]. This commercial device, however, has a memory depth of up to 2 Gpts with a sampling rate of 256 GS/s, and can thus record 7.8 ms traces. The jitter numbers, integrated here over a 100 Hz to 10 MHz frequency range, correspond to measurement times in the order of only 500 μs (e.g. equating Eqs. (4) and (5) in [11]). Rescaling them to a 7.8 ms measurement time, we obtain an ENOB of 5.4 for the all-electrical ADC estimate, in line with available commercial devices, and 7.4 for the optically enabled, four-slice ADC described above. Moreover, harmonic distortion and other sources of noise, that would occur in a practical implementation, have not been considered here and would further reduce the ENOB.

In conclusion, the aperture jitter of the MLL used as a spectral reference has been identified as the fundamental limitation in the noise performance of an optically enabled, spectrally sliced ADC. The jitter of the electrical oscillator driving the electrical ADCs used for digitization in each of the slices also limits the optically enabled ADC, its impact can, however, be reduced by increasing the number of slices. The phase noise of the optical carrier and the correlated MLL phase noise shared among all its lines have been shown to play a secondary role, as they can be compensated by proper system design and signal processing. Analytical results have been numerically verified. A four-slice architecture with an overall bandwidth of 120 GHz is expected to improve the ENOB of the all-electrical state-of-the-art by 2 with further improvements achievable by increasing the number of slices.